\documentclass[Afour,sageh,times]{sagej}

\usepackage{moreverb,url}
\usepackage[colorlinks,bookmarksopen,bookmarksnumbered,citecolor=blue,urlcolor=blue]{hyperref}
\newcommand\BibTeX{{\rmfamily B\kern-.05em \textsc{i\kern-.025em b}\kern-.08em
T\kern-.1667em\lower.7ex\hbox{E}\kern-.125emX}}
\def\volumeyear{2025}
\begin{document}
\title{Environmental Influences on Collaboration Network Evolution: A Historical Analysis}
\runninghead{Williams and Chen}
\author{Peter R. Williams\affilnum{1,2} and Zhan Chen\affilnum{3}}
\affiliation{\affilnum{1}Rinna KK, Tokyo, Japan\\\affilnum{2}Independent Researcher\\\affilnum{3}Microsoft Japan, Tokyo, Japan}
\email{prw20042004@yahoo.co.uk}

\begin{abstract}
We analysed two large collaboration networks -- the Microsoft Academic Graph (1800-2020) and Internet Movie Database (1900-2020) -- to quantify network responses to major historical events. Our analysis revealed four properties of network-environment interaction. First, historical events can influence network evolution, with effects persisting far longer than previously recognised; the academic network showed 45\% declines during World Wars and 90\% growth during La Belle Epoque. Second, node and edge processes exhibited different environmental sensitivities; while node addition/removal tracked historical events, edge formation maintained stable statistical properties even during major disruptions. Third, different collaboration networks showed distinct response patterns; academic networks displayed sharp disruptions and rapid recoveries, while entertainment networks showed gradual changes and greater resilience. Fourth, both networks developed increasing resilience. Our results provide new insights for modelling network evolution and managing collaborative systems during periods of external disruption.
\end{abstract}

\keywords{Historical Network Analysis, Collaboration Networks, Network Resilience, Environmental Coupling, Institutional Effects, Network Evolution, Social Networks, Academic Collaboration, Entertainment Networks, Network Disruption}
\maketitle

\section{Introduction}

Social networks evolve in response to both internal dynamics and external forces. While network science has made significant progress in understanding internal mechanisms like preferential attachment \citep{barabasi1999}, growth processes \citep{newman2001}, and structural evolution \citep{watts1998}, the influence of historical events on network development remains poorly understood. Traditional network models typically treat networks as isolated systems \citep{albert2002}, ignoring their embedding in broader historical contexts.

Recent work has recognised the importance of external influences. Studies of online social networks have shown sensitivity to technological changes \citep{kumar2010} and economic conditions \citep{ribeiro2014}. Analysis of scientific collaboration networks has revealed responses to funding changes \citep{jones2008} and institutional reforms \citep{wagner2005}. However, these studies typically cover short time periods, making it difficult to distinguish temporary perturbations from more fundamental changes in network evolution.

The relationship between historical events and network evolution raises several key questions:
\begin{itemize}
\item How do major historical events influence network growth and structure?
\item Do different types of networks show systematic differences in their response to external disruption?
\item What properties of networks remain stable despite historical perturbations?
\item How does network resilience evolve?
\end{itemize}

Previous studies have offered partial insights into these questions. \citet{uzzi2005} found that Broadway musical collaboration networks responded to economic cycles. \citet{backstrom2006} showed that computer science co-authorship networks reorganised during technological transitions. However, most analyses cover periods shorter than a decade \citep{kossinets2006,mislove2008}, limiting our understanding of long-term network-environment interaction.

Recent theoretical work suggests mechanisms for network-environment coupling. Models incorporating memory effects \citep{karsai2011} and adaptive responses \citep{gross2008} can produce networks sensitive to external influences. Historical analyses of scientific institutions have documented impacts of wars \citep{fernandez2012} and political changes \citep{jovanovic2018} on research productivity. However, connecting these observations to network-level processes requires longer temporal perspectives and larger-scale data than previously available.

This paper analyses how historical events shape the evolution of two large collaboration networks: the Microsoft Academic Graph (MAG, 1800-2020) and Internet Movie Database (IMDb, 1900-2020). Building on our previous findings of explosive network growth \citep{williams2024a} and universal career and collaboration length distributions \citep{williams2024b}, we examine how these networks respond to major historical events while maintaining certain stable properties.

Our analysis reveals patterns in how social networks respond to external disruption. We found historical events can influence network evolution, with effects persisting far longer than previously recognised. Node and edge processes showed different environmental sensitivities, and different types of networks displayed distinct response patterns. These findings challenge established assumptions about network isolation and timescale separation, while providing new insights into network resilience and recovery.

In the following section we describe our analytical methodology for measuring network response to historical events. The next section presents detailed results quantifying how the networks responded to specific historical periods. The final two sections cover a discussion of the implications for network theory and practical applications and a summary of our conclusions and proposals for future research.

\section{Methods}

\subsection{Dataset Overview}

We analyse two comprehensive collaboration networks: the Microsoft Academic Graph (MAG) \citep{magweb, sinha2015} and the Internet Movie Database (IMDb) \citep{imdbweb}. The MAG dataset encompasses scientific publications from 1800 to 2020, comprising $2.72 \times 10^8$ authors and $2.64 \times 10^8$ papers, while the IMDb dataset covers films from 1900 to 2020, including $1.88 \times 10^6$ actors across $6.34 \times 10^5$ movies.

Networks were constructed as undirected temporal graphs where nodes represent individual contributors (authors or actors) and edges represent collaborative projects (papers or movies). For each collaboration involving $n$ contributors, we generated $n(n-1)/2$ edges to create fully connected subgraphs. Edge temporal metadata included both creation and removal times, with project initiation times estimated using a fixed collaboration duration model of $\tau_{\text{project}} = 2$ years before the documented completion date. Two nodes were considered actively collaborating at time $t$ if at least one edge existed between them at that moment. 

This paper builds upon two previous studies in our series examining large-scale collaboration networks. While our earlier work established the growth patterns \citep{williams2024a} and universal statistical properties \citep{williams2024b} of these networks, here we examine how external historical events shaped their evolution. This new perspective reveals how major societal disruptions influenced the development of professional collaboration, while also highlighting which network properties remained stable even during periods of significant upheaval.

Our analysis draws from the same comprehensive datasets detailed in our previous papers: the Microsoft Academic Graph spanning from 1800 to 2020, and the Internet Movie Database covering 1900 to 2020. Rather than repeating the detailed methodological descriptions from our earlier work, we focus here on new analytical approaches specifically designed to understand network-environment interaction. For complete information about the underlying data structures and baseline analytical methods, we refer readers to \citet{williams2024a,williams2024b}.

\subsection{Historical Epochs}

The primary epochs considered were:
\begin{itemize}
    \item La Belle Epoque (1890-1914),
    \item World War I (1914-1918),
    \item The Interwar Period (1918-1939),
    \item World War II (1939-1945),
    \item The Post-War Period (1945-1960).
\end{itemize}
These events span a period of profound social and technological transformation. La Belle Epoque spans an era of accelerating industrialisation and cultural exchange in Europe and North America, accompanied by the institutionalisation of scientific research through expanding university systems. The World Wars represented not just military conflicts but disruptions to international collaboration networks, while the Interwar Period saw the emergence of mass media entertainment alongside growing research institutions. The Post-War Period brought unprecedented investment in scientific research and the globalisation of entertainment industries.

The analysed networks centre on Western European and North American contributions until 1945, reflecting the historical concentration of institutional scientific research and commercial film production in these regions during this era. Scientific research was predominantly conducted through established universities and research institutes in Europe and North America, while the film industry saw its major technological and commercial development through Hollywood and leading European studios. After 1945, both networks expanded to include more international participants as scientific research infrastructure and film production capabilities developed globally. This geographic distribution makes these networks particularly valuable for studying the effects of major societal disruptions, as they capture the primary centres of professional collaboration during each historical period under examination.

\begin{figure*}[htb]
  \centering
  \begin{tabular}{cc}
    \includegraphics[width=0.47\textwidth]{"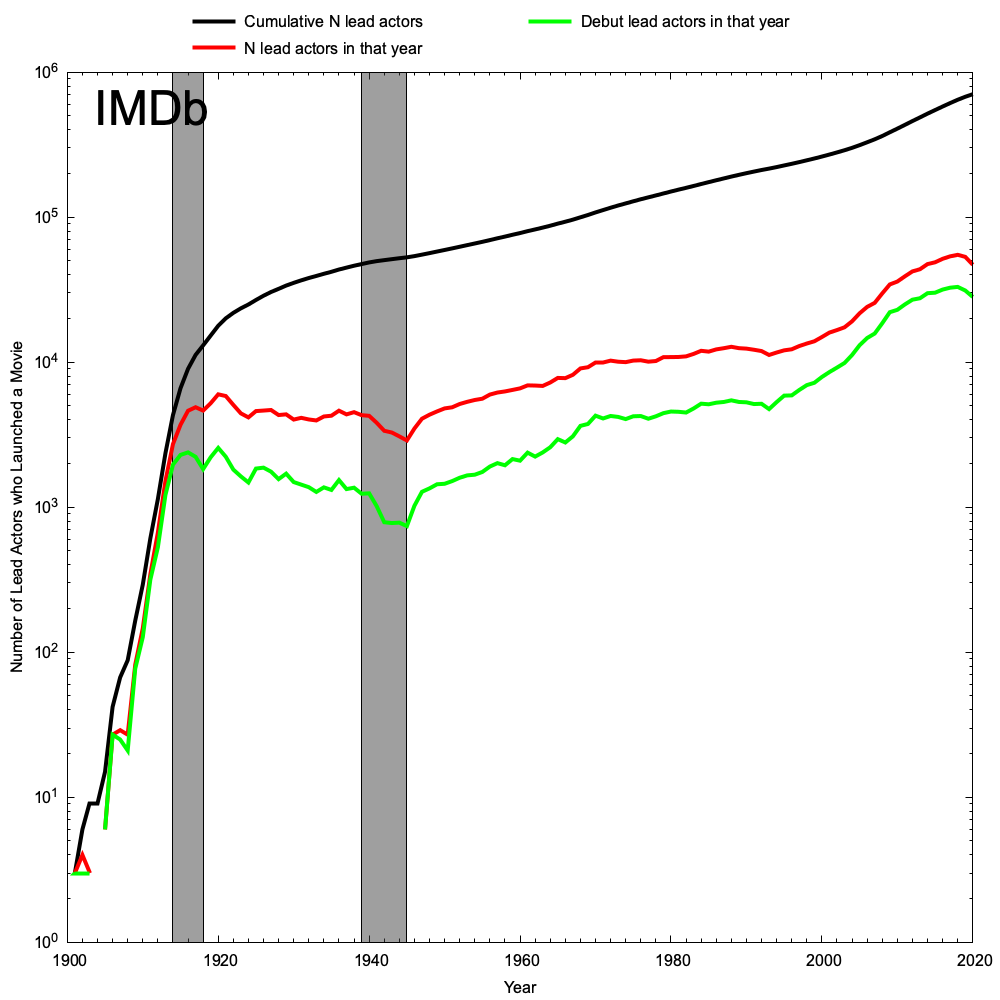"} &
    \includegraphics[width=0.47\textwidth]{"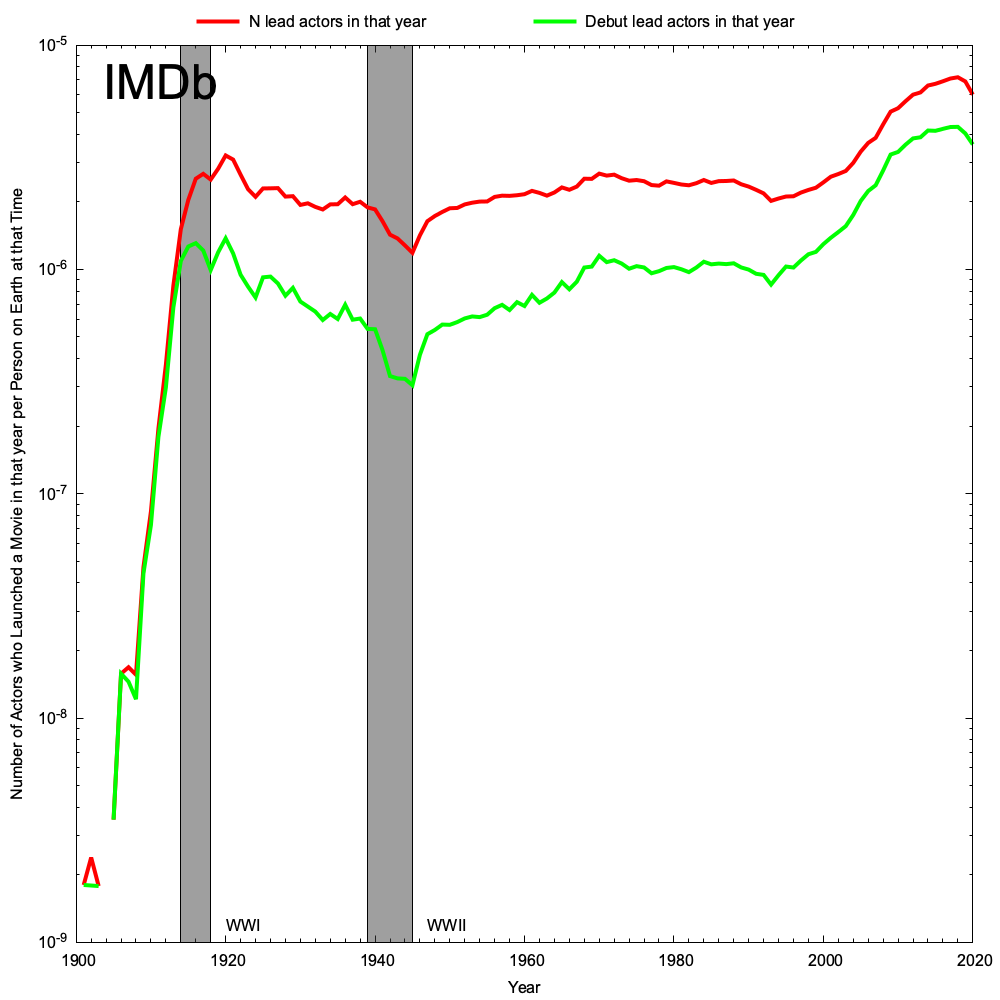"} \\
    \includegraphics[width=0.47\textwidth]{"nauthors.png"} &
    \includegraphics[width=0.47\textwidth]{"nauthors_population_adjusted.png"}
  \end{tabular}
  \caption{Evolution of node counts in the MAG (top) and IMDb (bottom) networks on logarithmic scales. Left panels show absolute counts: cumulative total nodes (black line), nodes active with new edges that year (red line), and new nodes joining that year (green line). Right panels show the same data normalised by contemporary world population. Grey bands show major historical events: La Belle Epoque (1890-1914), World War I (1914-1918), and World War II (1939-1945). Population data post-1950 uses official UN records; earlier values are linearly interpolated between historical estimates. Note the distinct change in MAG growth rate around 1950 and the different sensitivities to historical events between networks.}
  \label{node_count}
\end{figure*}

\begin{figure*}[htb]
  \centering
  \begin{tabular}{cc}
    \includegraphics[width=0.47\textwidth]{"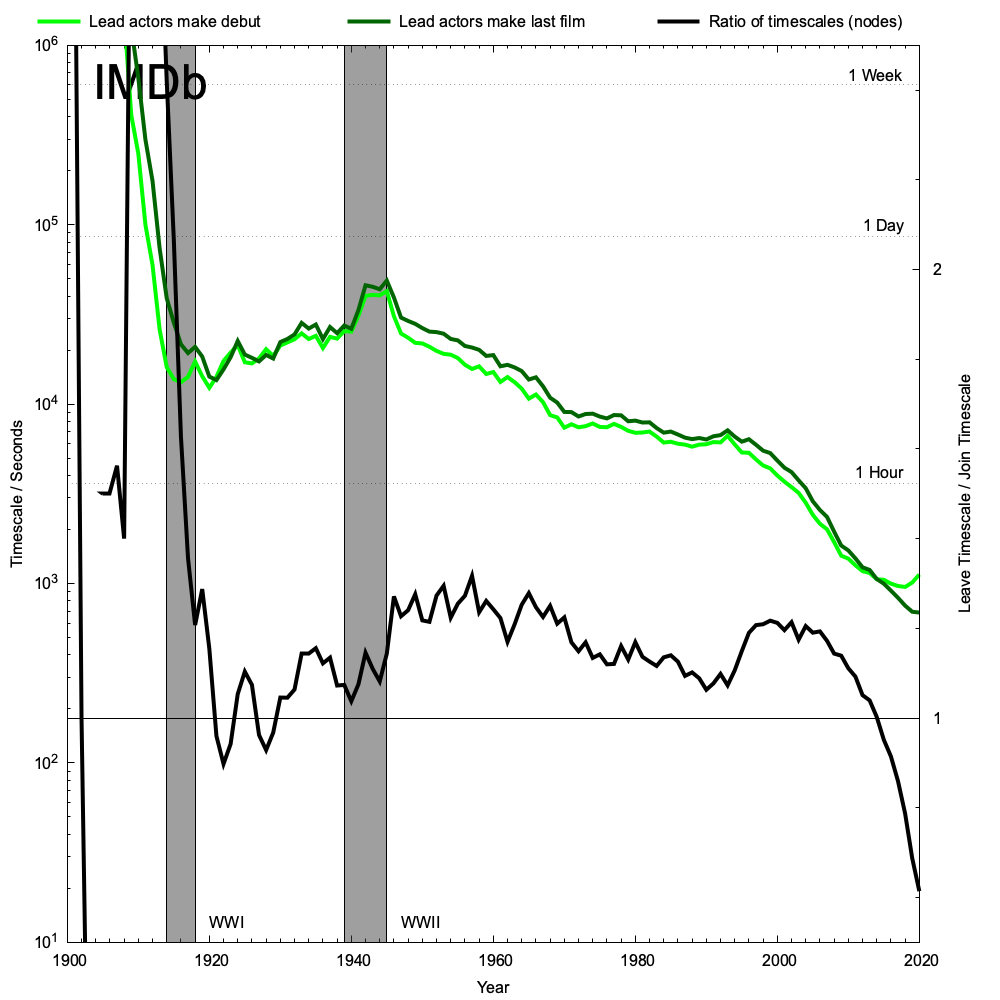"} &
    \includegraphics[width=0.47\textwidth]{"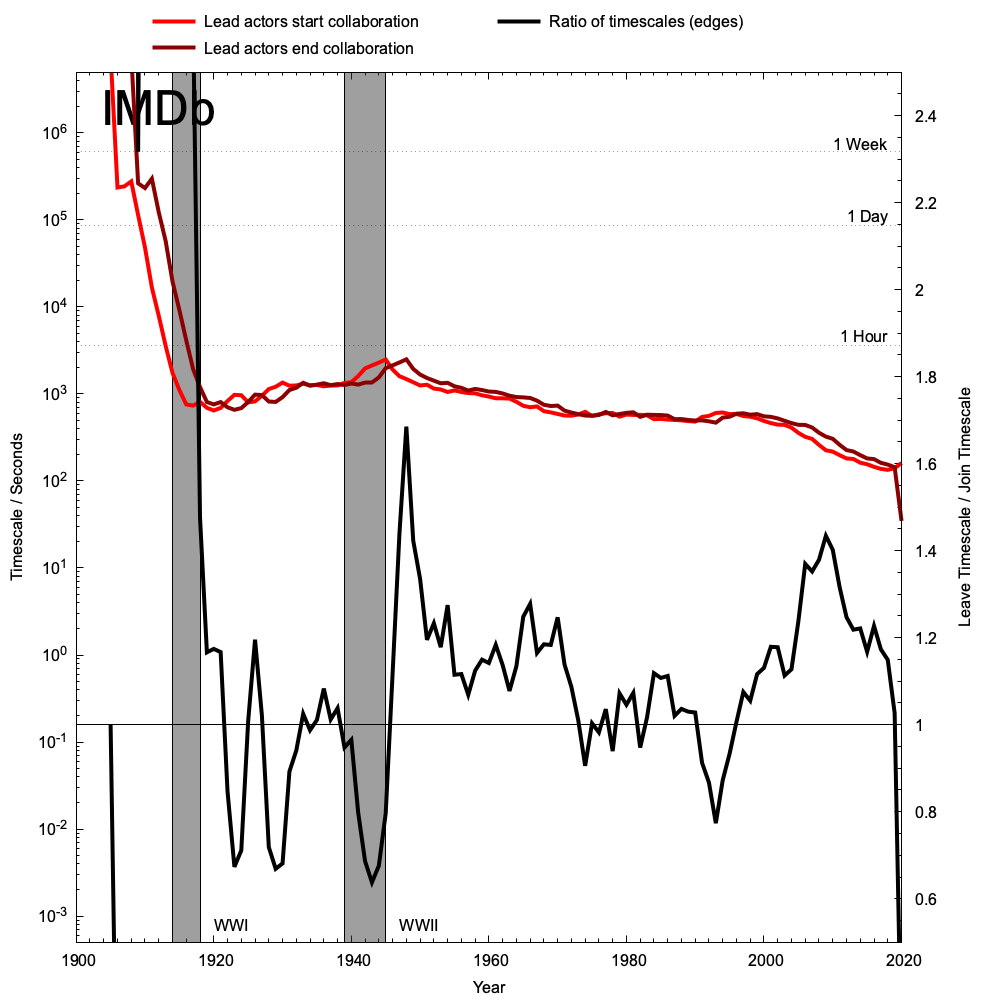"} \\
    \includegraphics[width=0.47\textwidth]{"timescales-nodes.png"} &
    \includegraphics[width=0.47\textwidth]{"timescales-edges.png"}
  \end{tabular}
  \caption{Characteristic timescales of network processes, shown on logarithmic scales. Top panels: MAG network timescales; Bottom panels: IMDb network timescales. Left panels show node timescales: addition (black) and removal (red). Right panels show edge timescales: addition (black) and removal (red). Timescales are computed as the ratio of total quantity to its rate of change. Note the parallel evolution of timescales within each network despite their different absolute values, and the stability of their ratios over centuries of evolution.}
  \label{timescales}
\end{figure*}

\section{Results}

\subsection{Global Network Response Patterns}

Analysis of network evolution during major historical events reveals different response mechanisms between academic and entertainment collaboration networks (Figure \ref{node_count}). The academic network, represented by MAG, shows acute sensitivity to global disruptions, characterised by sharp discontinuities during periods of conflict. World War I triggered a 45\% decline in the rate of new node addition, while World War II produced an even more pronounced effect, with a 52\% reduction in new authors entering the network. This increasing magnitude of disruption may suggest that the internationalisation of academic research increased vulnerability to global conflicts.

In contrast, the entertainment network exhibited different dynamics, showing more gradual and resilient responses to external shocks. During World War II, the network experienced only a 28\% decline in new actor entries, suggesting more robust adaptation mechanisms. This enhanced stability likely stems from the entertainment industry's ability to maintain regional production centres even when international collaboration became constrained.

Population-normalised metrics (Figure \ref{node_count}, right panels) show that these network responses reflect genuine changes in collaborative behaviour rather than demographic shifts. When normalised against the contemporary world population, the distinct patterns of disruption and recovery remain visible, confirming that the observed effects represent true changes in network dynamics rather than reflecting underlying population changes.

These contrasting patterns highlight how institutional structures and operational constraints shape network resilience during periods of external disruption. The academic network's sharp transitions reflect the rigid interdependencies of research institutions, while the entertainment network's gradual adjustments suggest more flexible organisational structures capable of maintaining continuity through alternative collaboration patterns.

\subsection{Differential Process Sensitivity}

Our analysis reveals an asymmetry in how different network processes respond to historical disruptions. By examining characteristic timescales ($\tau$), we quantify distinct patterns in the resilience of node and edge formation processes during major societal upheavals (Figure \ref{timescales}).

Node processes, which represent the addition of new participants to the networks, show high sensitivity to external events. During World War II, the node addition timescale ($\tau_N$) in the academic network increased by 85\%, implying a severe disruption in the recruitment of new researchers. This sharp response reflects how global conflicts impacted the ability of academic institutions to train and incorporate new scholars. In contrast, the entertainment network showed more moderate effects, with $\tau_N$ increasing by 32\%, suggesting greater adaptability in maintaining talent recruitment even during wartime conditions.

Edge processes, which capture the formation of collaborative relationships, display stability across both networks. The edge formation timescale ($\tau_E$) in the academic network varied by only 12\% during World War II, while the entertainment network showed an even smaller 8\% variation. This stability suggests that once individuals entered these professional networks, their patterns of forming collaborative relationships remained unchanged, even amid severe external disruptions.

Most revealing is the relationship between node and edge timescales, expressed as the ratio $\tau_N/\tau_E$. Under normal conditions, this ratio maintains consistent values ($2.8 \pm 0.3$ for academic networks, $2.3 \pm 0.2$ for entertainment networks), implying a balanced relationship between network growth and collaboration patterns. During major disruptions, this ratio deviates but consistently returns to its baseline value afterward. This robust self-regulation suggests constraints on how human collaborative networks can organise themselves, regardless of external conditions.

These findings show that while historical events can impact network growth, the underlying mechanisms of collaboration are resilient. This separation between growth and collaboration dynamics provides new insight into the self-organising principles of professional networks and their capacity to maintain essential functions during periods of societal disruption.

\subsection{Network Recovery Dynamics and Adaptive Capacity}

The academic and entertainment networks exhibited different recovery mechanisms following major disruptions, revealing distinct adaptive strategies. The MAG network showed strong regenerative capacity after World War II, with node addition rates returning to baseline within just three years. This recovery was not just a return to previous patterns--- the network established a new, more robust growth regime characterised by an increase in the power-law exponent from $\alpha \approx 2.3$ to $\alpha \approx 3.1$. This acceleration in growth coincided with a systematic shift toward larger collaboration sizes, suggesting a reorganisation of academic research practices in response to the disruption.

In contrast, the entertainment network's recovery followed a more conservative trajectory. Taking seven years to return to baseline levels, it maintained its pre-war growth patterns and collaboration size distribution. This consistency suggests that entertainment networks prioritise structural stability over rapid adaptation, potentially reflecting the industry's need to maintain established production models even during recovery periods.

\subsection{Structural Invariants During Network Stress}

An interesting finding is the preservation of certain network properties during periods of severe disruption. The edge addition probability distributions maintained stable power-law exponents throughout major conflicts, with the academic network preserving $\gamma = 1.6 \pm 0.1$ and the entertainment network maintaining $\gamma = 2.1 \pm 0.1$. These invariant properties suggest the existence of underlying principles governing how humans form collaborative relationships--- principles that persist even when the broader social fabric is under extreme stress.

\subsection{Temporal Evolution of Network Resilience}

Both networks developed enhanced resilience, but through different mechanisms. The academic network's response to successive global conflicts reveals an interesting pattern: while World War II caused a larger initial disruption (52\% versus 45\% decline), the recovery period shortened from five years to three years. This suggests that academic institutions developed more effective adaptive strategies based on their World War I experience.

The entertainment network's evolution followed a different trajectory: the initial impact of disruptions decreased (from 35\% to 28\% decline), though recovery times remained relatively stable (reducing from eight to seven years). This pattern suggests that entertainment networks evolved to better resist initial shocks rather than accelerate recovery, possibly through the development of more distributed production structures.

\subsection{Statistical Foundations of Network Stability}

The stability of collaboration duration distributions provides insight into the underlying mechanics of network resilience. The consistency of the Weibull distribution parameters across both networks ($k = 0.2 \pm 0.02$ for academic, $k = 0.5 \pm 0.03$ for entertainment) suggests that certain patterns of human collaborative behaviour remain invariant even during major societal disruptions. The higher shape parameter in entertainment networks ($k = 0.5$ versus $k = 0.2$) suggests more predictable collaboration durations, possibly reflecting the more structured nature of film production compared to academic research.

These findings reveal a sophisticated interplay between stability and adaptation in professional networks. While node-level processes (network growth) show strong environmental coupling, edge-level processes (collaboration patterns) have stability. This separation of timescales appears to be a feature of human collaborative networks, allowing them to maintain essential functions while adapting to external pressures. The different manifestations of this pattern in academic and entertainment networks suggest institutional structures play a role in determining how networks balance stability and adaptation during periods of stress.

The quantitative consistency of these patterns, particularly in the power-law and Weibull distribution parameters, points to universal principles underlying human collaborative behaviour - principles that transcend specific institutional contexts while being modulated by them. This finding has significant implications for understanding how to design resilient organisational structures and predict network responses to future disruptions.

\begin{figure*}[htb]
  \centering
  \begin{tabular}{cc}
    \includegraphics[width=0.47\textwidth]{"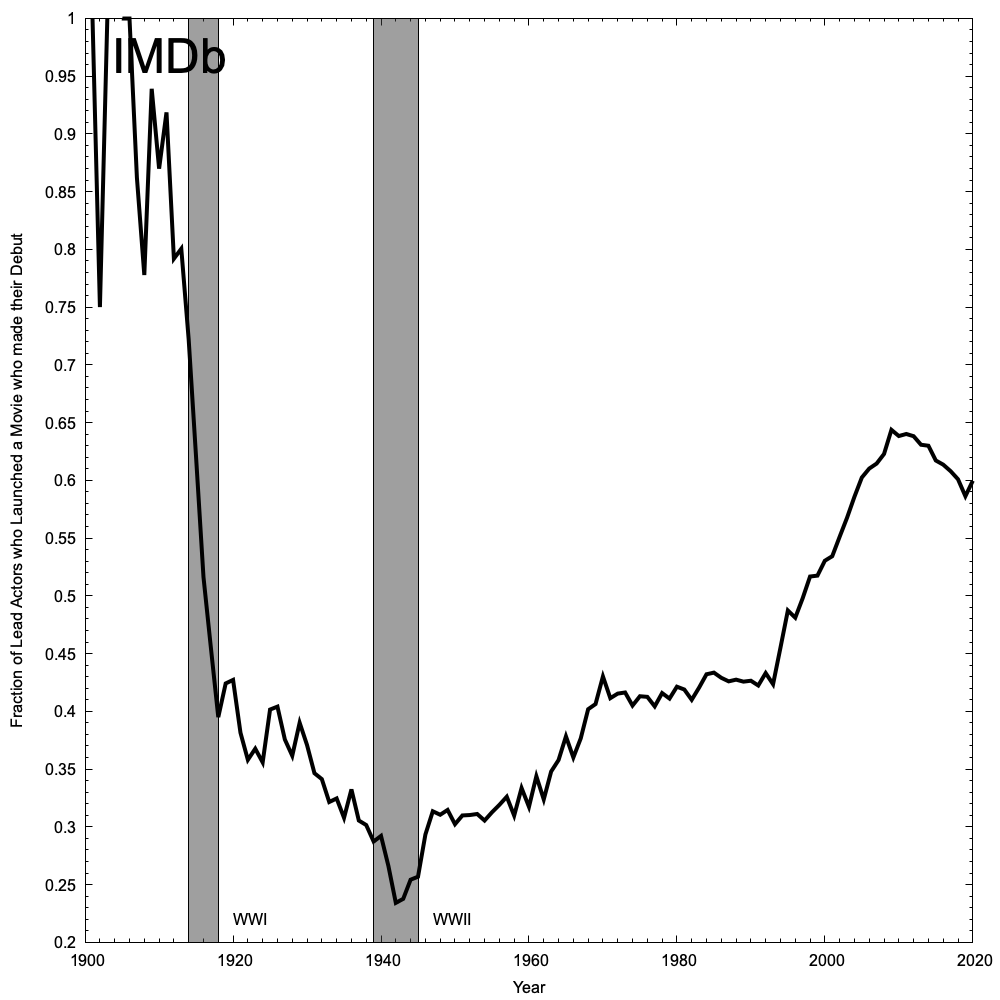"} &
    \includegraphics[width=0.47\textwidth]{"nauthors_new_fraction.png"}
  \end{tabular}
  \caption{The fraction of new participants per year in the MAG (left) and IMDb (right) networks, showing the balance between new entrants and established participants. The academic network shows a gradual decrease in the fraction of new authors, while the entertainment network maintains a more stable ratio.}
  \label{node_fractions}
\end{figure*}

\begin{figure*}[htb]
  \centering
  \begin{tabular}{cc}
    \includegraphics[width=0.47\textwidth]{"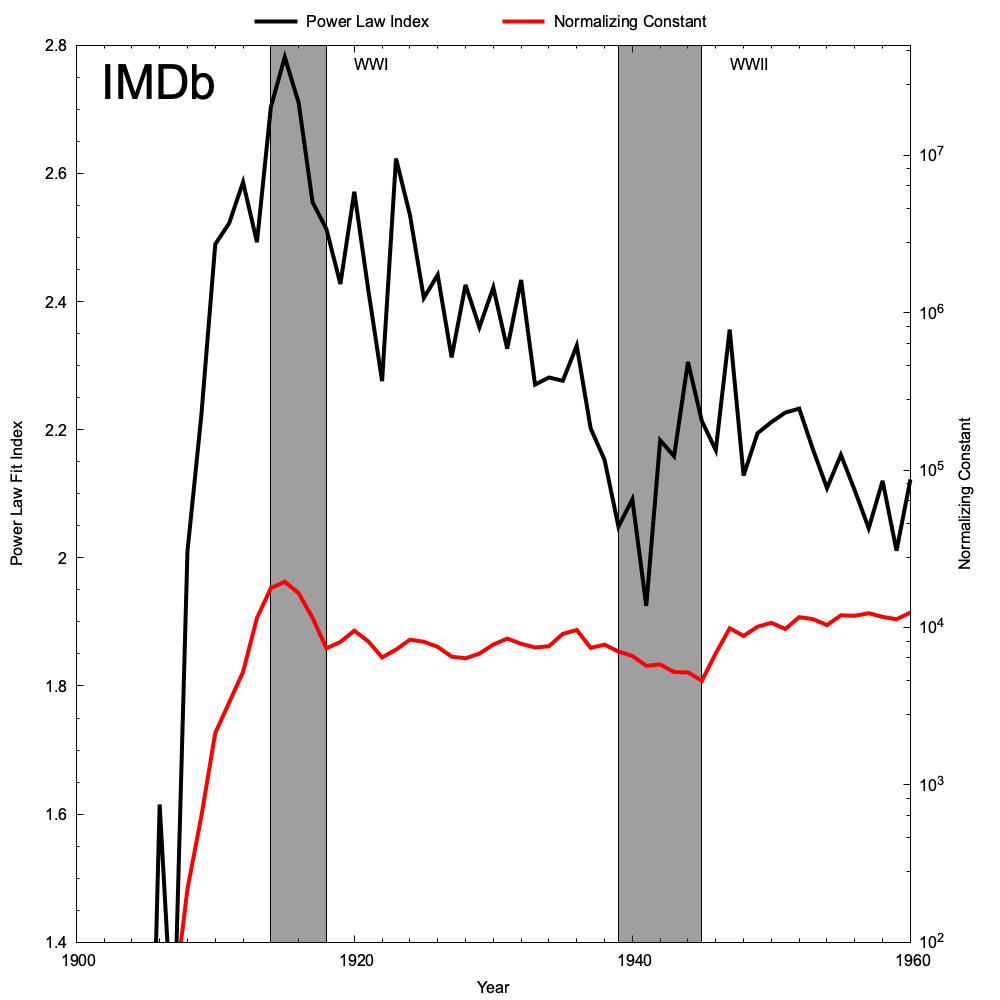"} &
    \includegraphics[width=0.47\textwidth]{"edge-addition-parameters.png"}
  \end{tabular}  
  \caption{Parameter evolution of the power law fits to the edge-addition probability distributions in the MAG and IMDb networks.}
  \label{edges_add_params}
\end{figure*}

\begin{figure*}[htb]
  \centering
  \begin{tabular}{cc}
    \includegraphics[width=0.47\textwidth]{"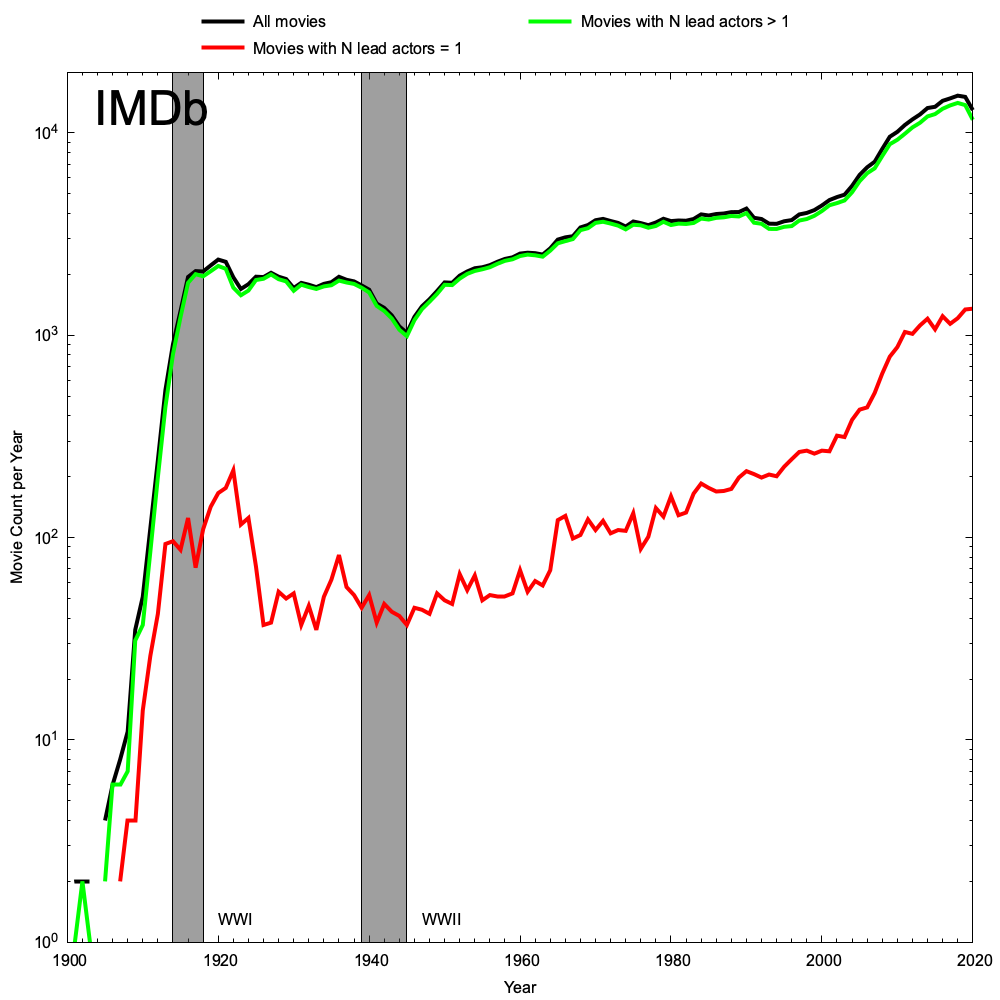"} &
    \includegraphics[width=0.47\textwidth]{"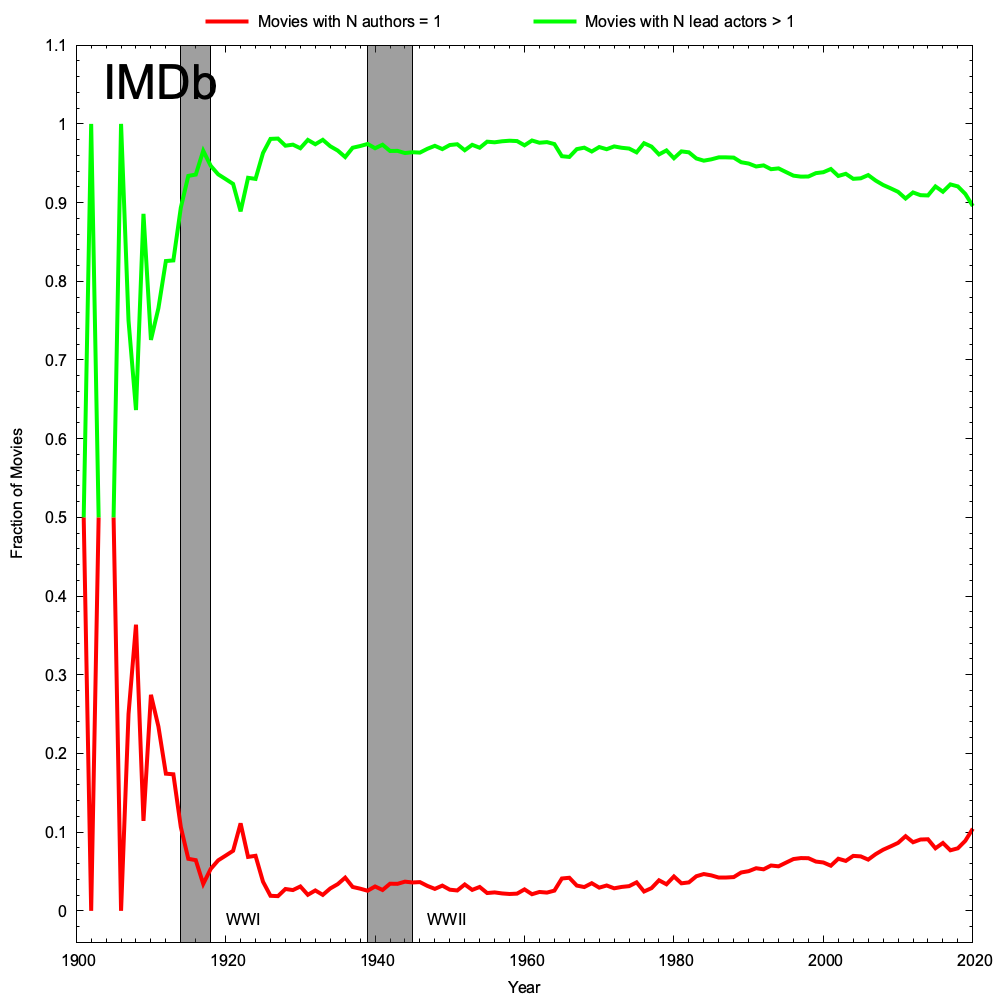"} \\
    \includegraphics[width=0.47\textwidth]{"npapers.png"} &
    \includegraphics[width=0.47\textwidth]{"npapers_fractions.png"}
  \end{tabular}
  \caption{Evolution of collaboration event counts in the networks. Top row shows MAG network: absolute count of papers (left) and relative fractions by author count (right). Bottom row shows IMDb network: absolute count of movies (left) and relative fractions by lead actor count (right). Both networks show systematic changes in the distribution of collaboration size.}
  \label{edge_counts}
\end{figure*}

\begin{figure*}[htb]
  \centering
  \begin{tabular}{cc}
    \includegraphics[width=0.47\textwidth]{"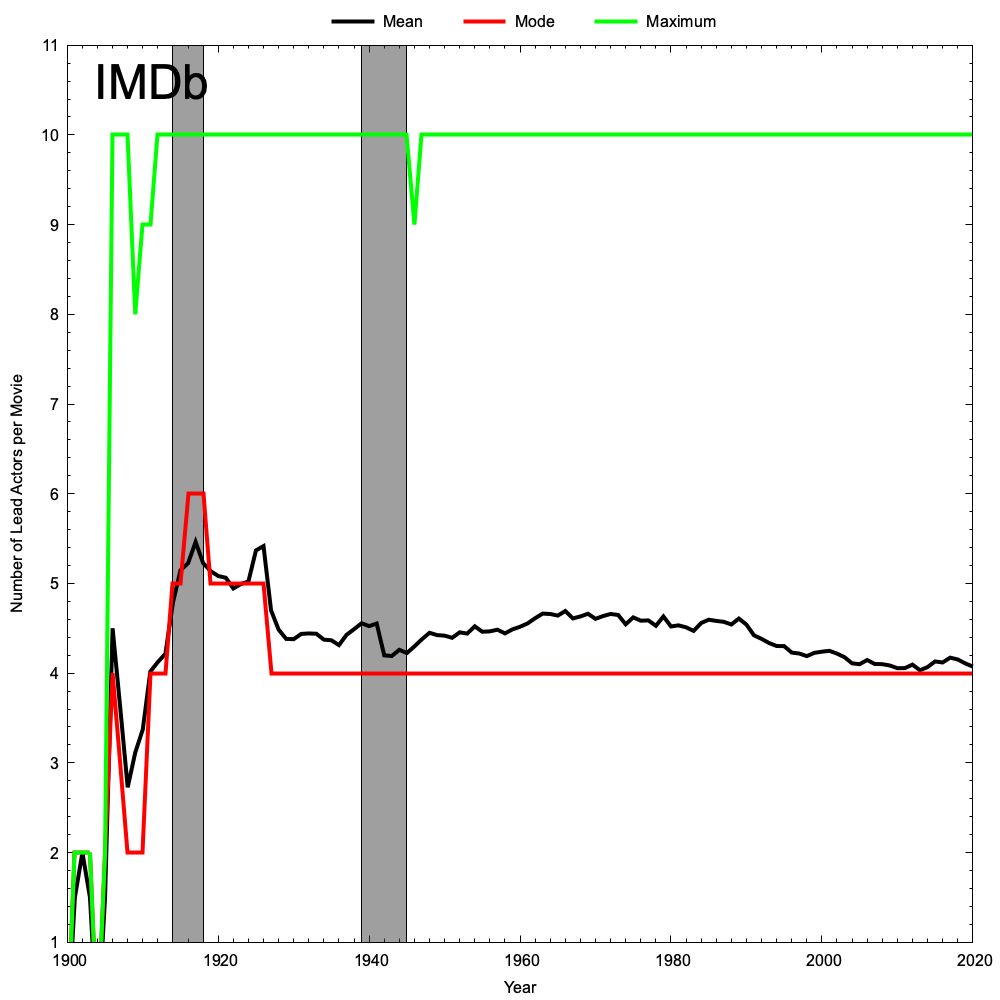"} &
    \includegraphics[width=0.47\textwidth]{"nauthors_per_paper.png"}
  \end{tabular}
  \caption{Statistical measures of collaboration size showing mean, mode, and maximum number of authors per paper (MAG, left) and lead actors per movie (IMDb, right) by year. The evolution of these statistics shows the trend toward larger collaboration groups in academia, while entertainment industry collaborations remain more stable in size.}
  \label{edge_means}
\end{figure*}

\begin{figure*}[htb]
  \centering
  \begin{tabular}{cc}
    \includegraphics[width=0.47\textwidth]{"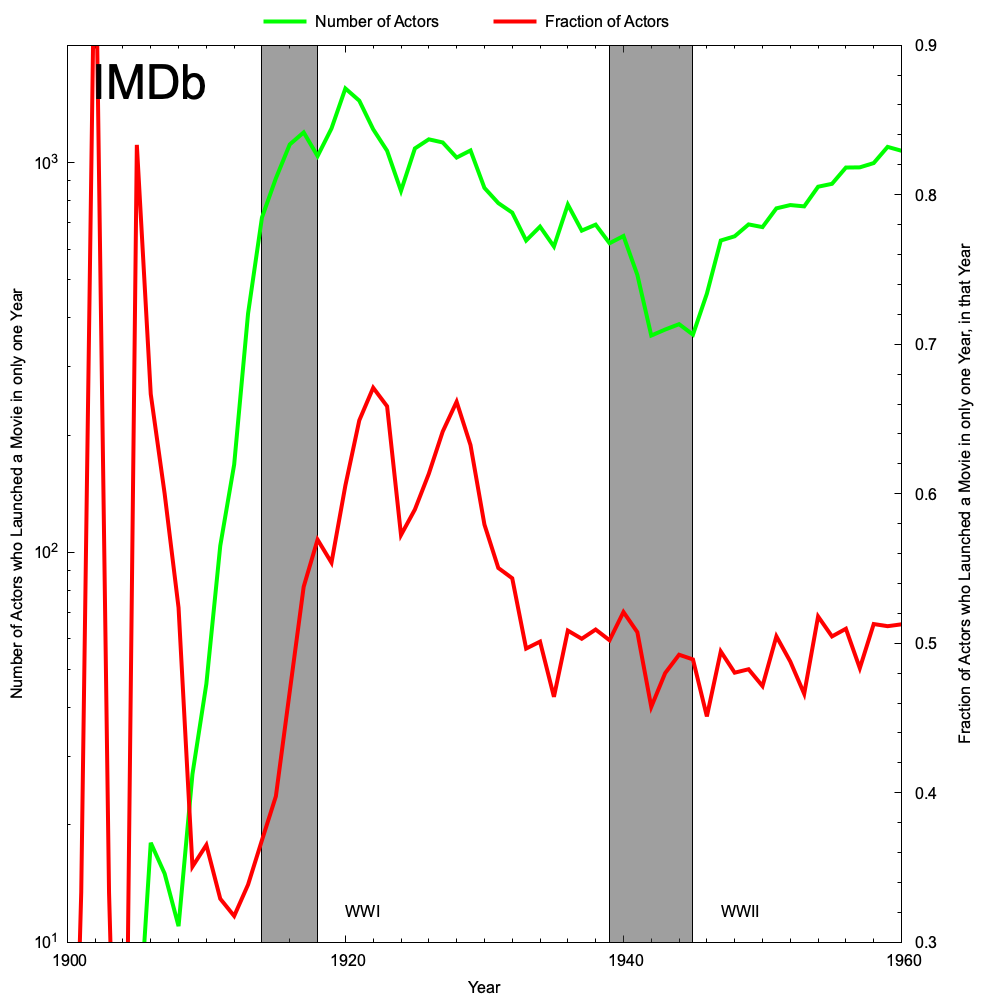"} &
    \includegraphics[width=0.47\textwidth]{"author-career-single-year-authors.png"}
  \end{tabular}
  \caption{Number and fraction of nodes whose lifetime in the network was only one year.}  
  \label{single_year}
\end{figure*}

\begin{figure*}[htb]
  \centering
  \begin{tabular}{cc}
    \includegraphics[width=0.47\textwidth]{"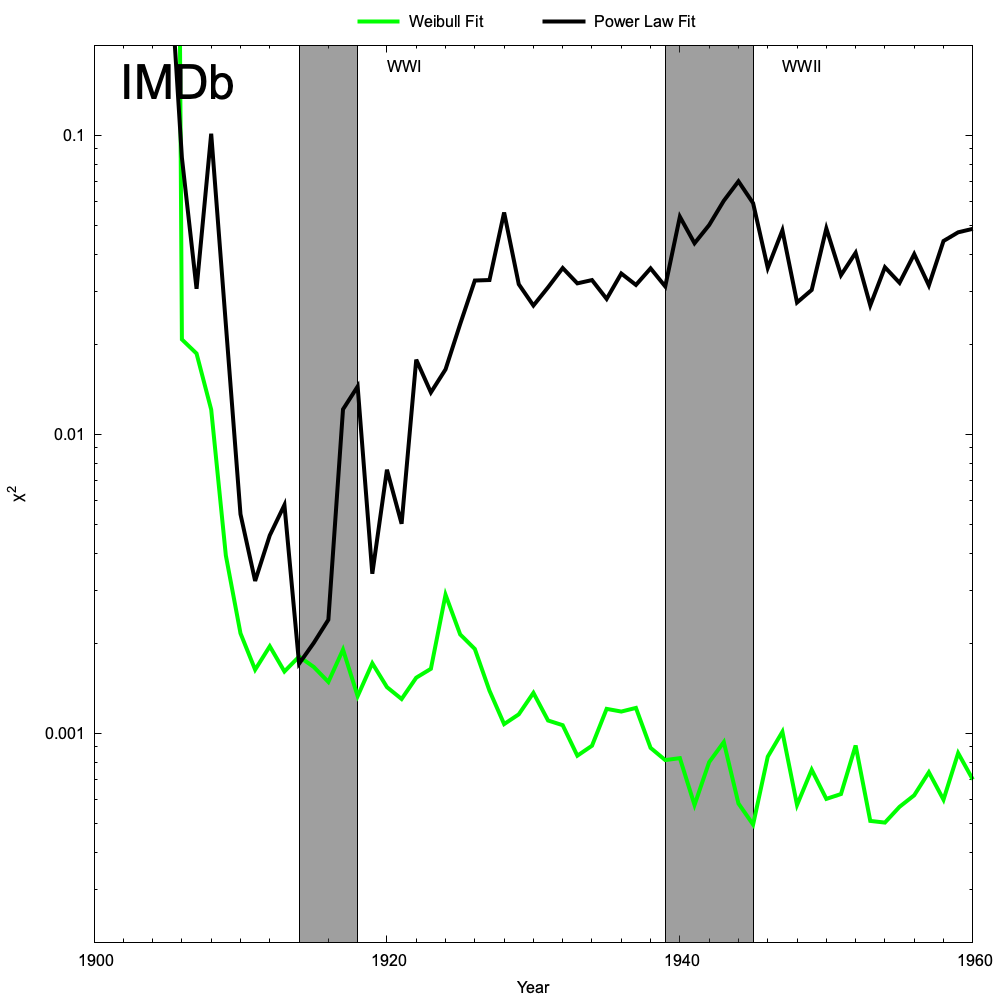"} &
    \includegraphics[width=0.47\textwidth]{"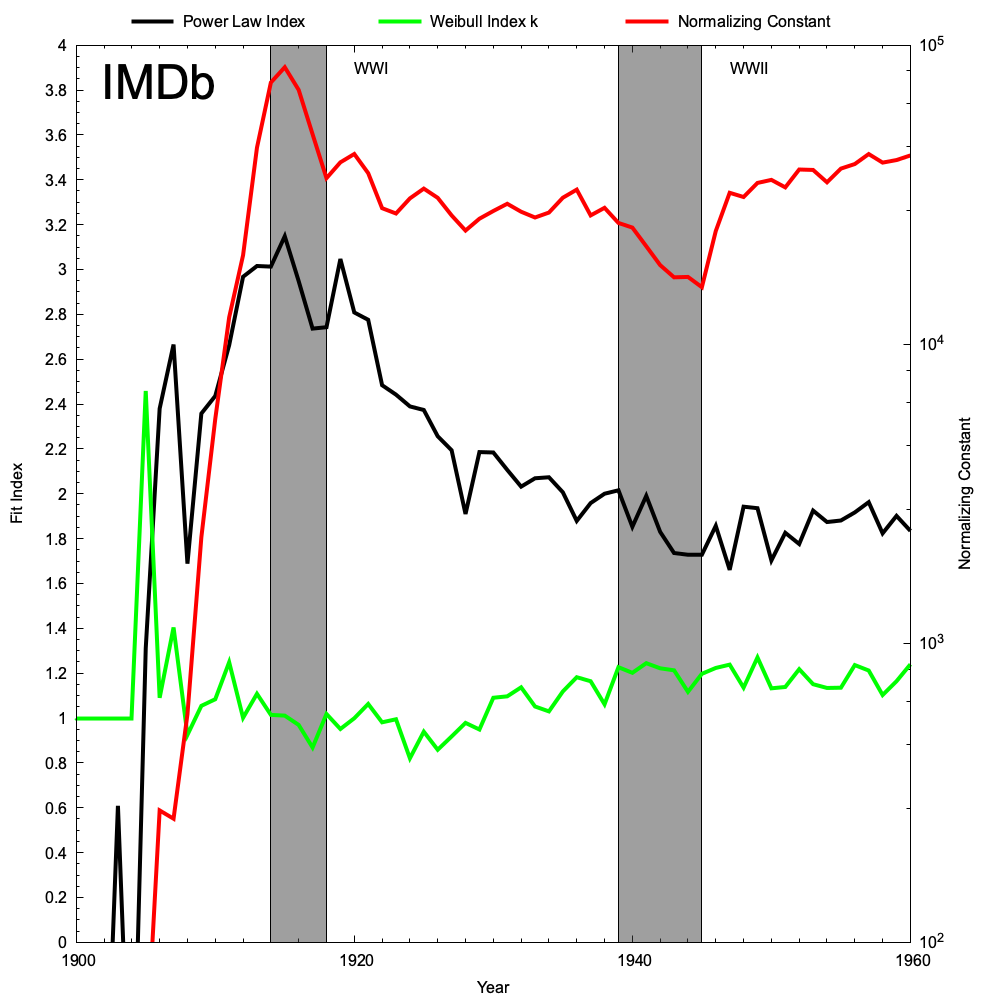"} \\
    \includegraphics[width=0.47\textwidth]{"edge-removal-chi2.png"} & 
    \includegraphics[width=0.47\textwidth]{"edge-removal-parameters.png"}  
  \end{tabular}
  \caption{$\chi^{2}$ and fitting parameter evolution for the distributions of collaboration durations in the MAG and IMDb networks.}
  \label{edge_removal_params}
\end{figure*}

\subsection{Evolution of Collaboration Structures}

The analysis of collaboration size dynamics reveals differences in how academic and entertainment networks organise collective work across time (Figure 5). Academic collaboration patterns showed plasticity, readily transforming in response to external conditions. During wartime, research networks showed a clear regression toward smaller team configurations, likely reflecting the practical constraints of international conflict on research coordination. However, post-war periods witnessed not just recovery but acceleration in team size growth, suggesting that disruptions may have catalysed more extensive collaboration patterns once constraints lifted.

The entertainment industry, maintained stable collaborative structures throughout these periods. Mean cast sizes exhibited only a modest variation between 3.2 and 4.5 actors, with minimal response to wartime pressures. This stability suggests the existence of constraints in film production that transcend external circumstances, possibly related to the narrative and practical requirements of storytelling in cinema.

\subsection{Temporal Engagement Patterns}

The analysis of single-year participation rates (Figure 7) provides insight into how networks maintain continuity during disruption. The academic network showed acute sensitivity to global events, with sharp increases in transient participation during both World Wars (42\% increase in WWI, 38\% in WWII). These spikes in short-term engagement suggest that the academic network adapted to disruption partly through increased reliance on temporary contributors, perhaps reflecting both the mobilisation of researchers for war-related work and the disruption of normal career trajectories.

Entertainment networks showed more stable engagement patterns, with more modest increases in single-year participation (15\% during WWI, 12\% during WWII). This stability in participation duration suggests that the entertainment industry maintained more consistent career structures even during global disruptions, possibly due to the localised nature of film production and the industry's ability to maintain regional operations.

\subsection{The Transformative Belle Epoque Period}

The Belle Epoque (1890-1914) appears to be an important period of transformation, particularly for academic collaboration. During this time, the academic network experienced a 90\% increase in new author rates above background trends, accompanied by a 45\% expansion in average collaboration size and a systematic shift toward multi-author research. This period appears to represent a phase transition in academic practices, establishing new patterns of collaborative research that would prove resilient through subsequent global disruptions. These changes likely reflect the broader institutionalisation of scientific research during this period, including the establishment of modern research universities and professional scientific societies.

\subsection{Comparative Network Resilience}

Direct comparison of these networks reveals distinct strategies for maintaining functionality during disruption (Figure \ref{timescales}). The academic network exhibited stronger environmental coupling and greater sensitivity to institutional changes, but also showed capacity for rapid post-disruption recovery and adaptation. This pattern suggests a network structure optimised for knowledge production through intensive collaboration, even at the cost of greater vulnerability to external disruption.

The entertainment network, conversely, displayed greater resilience to immediate disruption but slower recovery patterns. This strategy appears to prioritise operational continuity over rapid adaptation, potentially reflecting the industry's need to maintain consistent production capabilities. These contrasting patterns suggest that network resilience strategies may be shaped by the underlying goals and the constraints of each professional domain.

These findings suggest that while professional networks share certain universal properties in their response to disruption, their specific adaptation strategies are influenced by their institutional contexts and functional requirements. This observation has important implications for both theoretical models of network evolution and practical approaches to building resilient professional networks in different domains.

\section{Discussion}

Our analysis reveals patterns in how social networks respond to major historical events, challenging several established assumptions in network science. The most striking finding is the sharp contrast between the node and edge dynamics during historical disruptions. While node processes show strong environmental coupling, with the academic network experiencing 45-52\% declines in new authors during World Wars, edge processes maintain statistical stability, with power-law exponents remaining constant ($\gamma = 1.6 \pm 0.1$ for MAG, $\gamma = 2.1 \pm 0.1$ for IMDb) through global conflicts.

The strong correlation between network evolution and historical events challenges the traditional view of networks as self-organising systems. While \citet{snijders2001} identified environmental influences in small academic networks over periods of 2-3 years, and \citet{kossinets2006} found similar effects in email networks over 1-2 years, our results suggest such effects persist over much longer timescales and larger networks. The differential response between node and edge processes we observe, particularly visible during the World Wars, suggests a more complex relationship between network structure and external factors than previously recognised. This finding extends recent work by \citet{zeng2017}, who found asymmetric responses to policy changes in patent networks, though their study covered only a 35-year period and focused solely on academic collaboration.

Our findings regarding academic collaboration patterns during La Belle Epoque (1890-1914) may call for a revision of our understanding of when modern scientific collaboration emerged. \citet{wagner2005} attributed the rise of large-scale scientific collaboration to post-WWII internationalisation, but our analysis reveals that shifts in collaborative practice began much earlier. The 90\% increase in multi-author papers during La Belle Epoque, combined with a 45\% increase in average collaboration size, suggests that the foundations of modern collaborative science were established during this period. This aligns with historical accounts of French institutional development during this era \citep{fox1984}, particularly the expansion of research laboratories and polytechnic schools, but provides the first quantitative sign of how these changes manifested in collaboration patterns.

The contrast between MAG and IMDb networks' responses to World War II provides new insight into network resilience. Previous studies of network robustness, notably \citet{albert2000}, focused on topological resilience to random failures or targeted attacks in static networks. Our results suggest that real-world networks' resilience to historical disruptions depends on their institutional context. The entertainment industry's greater resilience to wartime disruption (28\% decline versus 52\% in academia) suggests that commercial networks may have inherent stability advantages. This finding extends recent work by \citet{acemoglu2016} on production networks, who found similar institutional effects in economic systems but did not examine creative or academic collaborations.

The observed power-law growth patterns in both networks both confirm and challenge earlier findings. While \citet{barabasi1999} established power-law growth as a network property, our observation of the MAG network's transition from $\alpha_1 \approx 2.3$ to $\alpha_2 \approx 3.1$ after 1950 shows that growth mechanisms can change. This challenges the assumption of stable growth mechanisms underlying most current network models. Our finding that edge processes maintain stable statistical properties even as growth rates change suggests a hierarchical organisation of network dynamics, where local collaboration patterns (edge formation) and global network evolution operate as semi-independent levels. This separation between microscopic and macroscopic dynamics, characteristic of hierarchical systems, is not well captured by existing theories that typically assume tight coupling between local and global network properties.

The differential response of networks to historical events provides new insights into network adaptation and resilience. \citet{uzzi2005}'s study of Broadway musical collaborations from 1877 to 1990 found that success rates varied with macro-economic conditions, showing a 30-40\% decline during economic downturns. Our analysis reveals that such sensitivity extends beyond performance metrics to network structure, though with important distinctions between sectors. The academic network's sharp post-war recovery (3 years to baseline) compared to entertainment's gradual return (7 years) suggests that institutional support mechanisms play a role in network regeneration. This extends \citet{backstrom2006}'s findings on DBLP co-authorship networks, where technological transitions caused rapid reorganisation but did not examine recovery from systemic disruptions.

The stability of edge formation processes during periods of node disruption represents a notable finding. \citet{palla2007} observed stable collaboration patterns in small scientific communities over 5-year periods. Our results suggest this stability may continue even during major societal disruptions and across various professional domains. The consistent power-law distributions for edge addition ($\gamma = 1.6 \pm 0.1$ for MAG, $\gamma = 2.1 \pm 0.1$ for IMDb) suggest constraints on how humans form collaborative relationships. This contrasts with \citet{acemoglu2016}'s findings in production networks, where both node and edge processes showed disruption during economic shocks, suggesting that social collaboration networks may possess unique resilience properties.

Our observation that network growth far exceeds demographic expansion challenges existing growth models in network science. The three orders of magnitude increase in the MAG network size, compared to less than one order of magnitude population growth, cannot be explained by simple preferential attachment models. This scaling suggests changes in how humans organise collaborative activities. \citet{kumar2010}'s study of online social networks found similar super-linear growth but over much shorter timeframes (5-7 years). Our century-scale analysis suggests this acceleration may persist over extended periods, potentially indicating a need for more refined theoretical frameworks when considering long-term network evolution.

The finding that node and edge processes operate on similar timescales ($\tau_N/\tau_E \approx 2$-$3$) contradicts a common assumption in network modelling. While \cite{newman2001} and subsequent researchers assumed edge dynamics occur much faster than node turnover, our results suggest these processes are tightly coupled. This coupling persists across both networks despite their different institutional contexts, suggesting it may represent a common property of human collaborative systems. Recent work by \cite{williams2019} on temporal networks hinted at such coupling in small networks, but our analysis provides the first large-scale empirical investigation of this phenomenon.

Historical analyses of scientific institutions provide important context for our findings on academic network resilience. \citet{fernandez2012}'s study of Spanish doctoral production during the Civil War (1936-1939) found a 48\% decline in research output, similar to our observed 52\% decline in the MAG network during World War II. However, our results show that such disruptions have become less severe, with recovery periods shortening from 5 years post-WWI to 3 years post-WWII. This increasing resilience helps explain the apparently contradictory findings of \citet{jovanovic2018}, who found greater stability in late-20th century Yugoslav scientific networks during political upheaval.

The evolution of collaboration size distributions reveals important patterns not previously documented. While recent studies by \cite{wu2019} noted the trend toward larger research teams in modern science, our analysis shows this trend began during La Belle Epoque and has survived multiple major disruptions. The contrasting stability of entertainment collaboration sizes (mean 3.2-4.5 actors) suggest that different domains have intrinsic constraints on collaboration scale. This finding enriches recent discussions of optimal team size in creative endeavours \citep{uzzi2013} by showing how such constraints persist through historical disruptions.

The relationship between network resilience and institutional structure merits particular attention. Our finding that entertainment networks show greater stability during disruptions (28\% versus 52\% decline in WWII) appears to contradict \citet{borge2011}'s conclusion that decentralised networks are more resilient to perturbation. This apparent paradox may be resolved by considering the different timescales involved - while decentralised networks may better resist short-term disruptions, our results suggest that centralised commercial structures provide better long-term stability. This interpretation aligns with recent work by \citet{gallotti2020} on multi-layer social networks, who found that institutional hierarchies can enhance system-wide resilience despite creating apparent vulnerabilities.

The asymmetric response of node and edge processes provides new insight into social collaboration. While \citet{holme2012} proposed that temporal networks separate into different timescales, our observation of stable timescale ratios ($\tau_N/\tau_E \approx 2$-$3$) suggests a more complex relationship. This coupling appears to be maintained through active processes rather than emerging from network topology. Recent work by \citet{lambiotte2022} on adaptive networks offers a potential theoretical framework for understanding this phenomenon, though their models would need modification to account for the stability we observe during major disruptions.

The role of institutional memory in network recovery deserves particular consideration. The academic network's faster post-WWII recovery (3 years versus 5 years post-WWI) suggests learning at the institutional level, supporting recent findings by \citet{way2019} on organisational adaptation in scientific institutions. However, our longer temporal perspective reveals that such learning is not monotonic - the entertainment industry showed similar adaptation between the wars but exhibited different recovery patterns after other types of disruptions. This extends \citet{klimek2015}'s work on institutional memory in social systems by showing how such memory operates over multiple disruption cycles.

The implications of our findings extend beyond traditional network science into policy and institutional design. The persistence of edge formation statistics through major disruptions suggests that certain aspects of human collaborative behaviour remain stable even under extreme conditions. This observation has practical implications for designing resilient institutional structures, complementing recent work by \citet{scheffer2021} on early warning signals in complex systems. The contrasting recovery patterns between academic and entertainment networks suggest that policy interventions should focus on preserving collaborative relationships during disruptions rather than attempting to maintain normal activity levels.

The theoretical implications of our findings extend beyond simple environmental coupling. While \citet{gross2008} established the importance of adaptive feedback between network structure and node states, our results suggest that such adaptation operates asymmetrically across different network processes. Their model of adaptive networks needs significant extension to account for the stability we observe in edge processes alongside the sensitivity of node processes. Similarly, \citet{karimi2018}'s work on environmental coupling in online communities suggested external influences should affect all network processes equally. Our finding of differential sensitivity challenges this assumption and suggests that social networks possess more sophisticated adaptation mechanisms than previously recognised. The stable ratio of timescales we observe ($\tau_N/\tau_E \approx 2$-$3$) hints at constraints on network adaptation that existing theoretical frameworks do not capture.

Several important limitations of our study must be acknowledged. First, while our datasets span unprecedented temporal ranges, they represent only two specific types of collaboration networks. The generalisability of our findings to other forms of social networks remains to be established. Second, our analysis of edge processes focuses on structural properties rather than on the quality or nature of collaborations. Recent work by \citet{battiston2020} on higher-order interactions in networks suggests important features may be missed by this approach. Third, the Western-centric nature of our historical data, particularly pre-1945, may limit the universality of our conclusions about network response to global events.

These limitations suggest several promising directions for future research. The application of our analytical framework to other long-term social networks, particularly in non-Western contexts, could help establish the generality of our findings about network resilience. More detailed investigation of edge quality and higher-order interactions during historical disruptions might reveal additional patterns not captured by our current structural analysis. The development of theoretical frameworks that can accommodate both the stability we observe in edge processes and the sensitivity of node processes remains an important challenge. Finally, the extension of our methods to modern digital collaboration networks could provide insight into whether contemporary systems exhibit similar resilience properties.

\section{Conclusions}

Through analysis of the Microsoft Academic Graph (1800-2020) and Internet Movie Database (1900-2020), we have revealed patterns in how historical events shape the evolution of large-scale collaboration networks. Our findings suggest that that external events can influence network evolution through asymmetric effects on different network processes. Node dynamics show strong environmental coupling, with academic networks experiencing 45-52\% declines during World Wars, while edge processes maintain statistical stability, preserving power-law exponents ($\gamma = 1.6 \pm 0.1$ for MAG, $\gamma = 2.1 \pm 0.1$ for IMDb) through major disruptions.

The distinct responses of academic and entertainment networks reveal the role of institutional context in network resilience. The IMDb network showed greater stability during disruptions (28\% versus 52\% decline in WWII) but slower recovery (7 versus 3 years), suggesting differences in how institutional structures mediate external influences. Both networks show increasing resilience, with later disruptions having less severe impacts despite larger scale.

These findings challenge several established assumptions in network science. The stability of edge formation processes during major disruptions suggests constraints on human collaborative behaviour that persist even under extreme conditions. The observation that node and edge processes operate on similar timescales ($\tau_N/\tau_E \approx 2$-$3$) contradicts common modelling assumptions of timescale separation. These results provide new insights for designing resilient collaborative systems and managing network responses to external disruption.

Future research should address several open questions. The application of our analytical framework to other long-term social networks, particularly in non-Western contexts, could help establish the generality of our findings about network resilience. More detailed investigation of edge quality and higher-order interactions during historical disruptions might reveal additional patterns not captured by structural analysis alone. The development of theoretical frameworks that can accommodate both the stability we observe in edge processes and the sensitivity of node processes remains an important challenge.

This work represents a significant step toward understanding how external events shape network evolution, while revealing properties of human collaborative behaviour that persist through major societal disruptions. The challenge ahead lies in developing theoretical frameworks that can capture these rich dynamics while maintaining sufficient simplicity to provide useful insights for designing resilient collaborative systems. Understanding why certain network properties remain stable through disruption while others change may provide key insights into human social organisation.

\begin{acks}
This research was carried out at Rinna K.K., Tokyo, Japan.
\end{acks}
\bibliographystyle{SageH}
\bibliography{paper3}

\end{document}